\documentclass[sigconf,10pt]{acmart}
\usepackage{graphicx} 
\usepackage{pifont} 
\usepackage{float}
\usepackage{multirow}
\usepackage{array}
\newcolumntype{C}[1]{>{\centering\arraybackslash}p{#1}}  
\usepackage{listings}
\usepackage{tikz}
\usetikzlibrary{arrows.meta}
\usetikzlibrary{automata,positioning}
\usetikzlibrary{calc}
\usetikzlibrary{arrows.meta,arrows,backgrounds,patterns,positioning,shadings, decorations.markings, shapes.multipart, shapes}
\usepackage{xcolor}    
\usepackage{fancyvrb}  
\usepackage[capitalise,nameinlink]{cleveref}
\crefname{algorithm}{Alg.}{Alg.}
\Crefname{algorithm}{Alg.}{Alg.}
\crefname{section}{\S\!}{\S\S\!}
\Crefname{section}{Sec.}{Sec.}
\crefname{figure}{Fig.}{Fig.}
\Crefname{figure}{Fig.}{Fig.}
\crefname{equation}{Eqn.}{Eqn.}
\Crefname{equation}{Eqn.}{Eqn.}
\crefname{listing}{Listing}{Listings}
\Crefname{listing}{Listing}{Listings}
\crefname{defn}{definition}{definitions}

\newcommand{\remove}[1]{}

\lstdefinestyle{PromptStyle}{
  basicstyle=\ttfamily\small,
  keywordstyle=\color{blue!80!black},
  stringstyle=\color{green!60!black},
  commentstyle=\color{gray},
  identifierstyle=\color{black},
  backgroundcolor=\color{gray!10},
  showstringspaces=false,
  breaklines=true,
  breakindent=0pt, 
  frame=single,
  rulecolor=\color{gray},
  frameround=tttt,
  language=, 
  morekeywords={},
  morestring=[b]",
}
\newcommand{\siva}[1]{\textcolor{brown}{[Siva: #1]}}

\newsavebox\codeg

\title{Extremal Testing for Network Software using LLMs}

\author{\large Rathin Singha$^\dagger$, 
Harry Qian$^\dagger$, 
Srinath Saikrishnan$^\dagger$, 
Tracy Zhao$^\dagger$, 
Ryan Beckett$^\ast$,
Siva Kesava Reddy Kakarla$^\ast$,
George Varghese$^\dagger$
\\
\textit{\normalfont{$^\dagger$UCLA} \hspace{5pt}}
\textit{\normalfont{$^\ast$Microsoft Research} \hspace{5pt}}
}

\date{June 2025}
\usepackage{url}
\begin{document}

\begin{abstract}
  
Physicists often {\em manually} consider extreme cases when testing a theory. In this paper, we show how to {\em automate} extremal testing of network
software using LLMs in two steps: first, ask the LLM to generate input constraints (e.g., DNS name length limits); then ask the LLM to generate tests that violate the constraints.  We demonstrate how easy this process is by generating extremal tests for HTTP, BGP and DNS implementations, each of which uncovered new bugs. We show how this methodology extends to centralized network software such as  shortest path algorithms, and how LLMs can generate filtering code to reject extremal input.  We propose using agentic AI to further automate extremal testing. LLM-generated extremal testing goes beyond an old technique in software testing called Boundary Value Analysis.
  
\end{abstract}

\maketitle

\section{Introduction}
Extremal testing has been used for centuries by physicists~\cite{physicslimiting} and mathematicians~\cite{mathlimiting}, who often test a theory or a theorem on limiting cases such as infinite mass or the Alexander horned sphere~\cite{alexandersphere}.  In this paper, we ask ourselves: {\em can we generate 
extremal test cases automatically for programs, in particular network softwares?} Like the simplex method~\cite{simplex}, extremal testing traverses the corners of an input space unlike testing based on symbolic execution (e.g., ~\cite{kakarla2022scale}, ~\cite{singha2024messi}), which is more analogous to interior point methods~\cite{roos}.

Consider the intuitive validity requirements for a DNS name like \texttt{cnn.com.} A DNS name is at most 253 characters, labels should use ASCII characters, and labels should be separated by at most one dot. Violating these constraints --- by sending a DNS resolver a name with more than 253 characters, adding non-ASCII characters to a name or putting two consecutive dots in a name -- generates tests that may break some implementations. These tests are not generated by classic symbolic testing methods that generate tests for all valid paths through an implementation; further, solvers like KLEE\cite{cadar2008klee} would be hard pressed to generate a long string using a SAT solver~\cite{DBLP:conf/cav/GaneshD07} because KLEE models strings as finite arrays of characters. Fuzz testing with random inputs can theoretically generate such cases, but the probability of finding the right extremal cases (say with two consecutive dots) is extremely low\cite{fuzzer1}.

Our paper reports on our early experience with a small set of students, most with no networking experience, who used LLMs (GPT-4o, Deep Research, Deepseek, Gemini) to generate extremal tests in two steps. First, ask the LLM to generate validity conditions; then in a separate step, generate tests that violate the validity conditions. \cref{sec:prompting_startegies} shows that two-step prompting is more effective than one-step prompting.  Our claim is that extremal testing is a useful complement to other testing techniques for network software like symbolic testing and fuzz testing, finding new bugs that the other methods do not cover.  Further, the effort and expertise to generate such tests using LLMs in our chain-of-thought prompting methodology is low compared to other methods.  We hope our paper suggests a subarea of network software testing that others will develop.


Extremal testing is similar to a classic software testing methodology called Boundary Value Analysis (BVA)~\cite{zhang2015boundary}.  In BVA, if an input $I$ has to lie within a range $X < I < Y$, then tests are generated close to the boundaries including $X, X+1, Y-1, Y$. Extremal testing goes beyond BVA by using LLMs to generate a richer variety of constraints; most work in BVA focuses on simple range constraints and does not use LLMs. In extremal testing we also focus only on invalid inputs (e.g, $X$, $Y$) and not on other boundary values (e.g., $X+1, Y-1$) to finesse the 
oracle problem~\cite{oracleproblem} of knowing the right answer for valid inputs.



Recently, the use of LLMs for software testing has become a vast topic.  See ~\cite{LLMTestingSurvey} for a survey of 102 papers. However, existing work uses LLM-generated tests to increase code coverage~\cite{FacebookLLMTesting} or to enhance fuzzing/mutation testing~\cite{LLMTestingSurvey}, not to generate invalid inputs. This is because this research is in a context where: {\bf a)} the LLM does not understand the intent of the particular software being tested (unlike in our case for say DNS or HTTP); and {\bf b)} the human likely also does not know the full specification for that software. Thus their goals are improving coverage or generating better mutants, as proxies for testing the spec.  They also often need to generate new tests as the code evolves~\cite{FacebookLLMTesting}, while our tests can be generated once up front. \cref{sec:relatedwork} has more details.

Proceeding, ~\cref{sec:dns} gives a quick example for DNS.  ~\cref{sec:http} and ~\cref{sec:bgp} describe extremal testing of HTTP and BGP, and ~\cref{sec:software} shows extremal tests for shortest path 
software, while ~\cref{sec:futurework} suggests future work including using agentic AI to further automate extremal testing. 

\section{DNS Extremal Testing}
\label{sec:dns}

Our extremal testing methodology proceeds in two steps, prompting for constraints, and then asking to generate tests.  For DNS, we used: 

\noindent{{\bf Prompt 1:} {\em Tell me some validity constraints for DNS domain names.}}

GPT 4o's response (formatted to fit) is:

{\em 1. Length Constraints:} Each label (the parts between dots) must be between 1 and 63 characters long. The full domain name (including dots and the root dot) must not exceed 253 characters. The root domain is represented by an empty label (i.e., the trailing dot), which may be omitted in practice.

{\em 2. Character Constraints:}
Allowed characters in each label: A–Z, a–z (case-insensitive), 0–9, Hyphen (-). Labels must not start or end with a hyphen. Labels must not contain any other characters (no spaces, underscores, special symbols, etc.).


{\em 3. No Leading/Trailing Dots}: While a fully qualified domain name (FQDN) technically ends with a dot (example.com.), DNS resolvers usually accept domain names without a trailing dot.

{\em 5. No Empty Labels:} Two dots in a row (like example..com) are not allowed. An empty label only appears at the end to represent the root (i.e., example.com.).

{\em 6. Punycode for Internationalized Domain Names (IDNs)}: For non-ASCII characters (like m\"{u}nich.com), the domain must be encoded in Punycode, which converts it to an ASCII-compatible form (e.g., xn--mnich-kva.com).

{\em 7. DNS Label Restrictions for Hostnames:} For names that map to hosts (e.g., A or AAAA records), labels must follow stricter rules: RFC 1123 allows labels to start with a digit, which RFC 952 previously disallowed.

Extremal testing proceeds with the following prompt: 


\noindent{{\bf Prompt 2}: {\em Please generate a set of invalid DNS names that violate one or more of the constraints above and can be used to test real DNS resolver implementations:}}

ChatGPT 4.0's response is (formatted for clarity and brevity):

\noindent{Here is a set of invalid DNS names that each violate one or more DNS validity constraints.} 

\noindent{{\em Test 1:} X.example.com, where X consists of 64 a's (invalid because first label is 64 characters).} 

\noindent{{\em Test 2:} Y.example.com, where Y consists of "a." repeated 130 times
(invalid because entire name is over 253 characters (counting dots and the root dot).}

\noindent{{\em Test 3:} exam!ple.com (invalid hostname because "!" is not allowed in DNS labels).}

\noindent{{\em Test 4:} test\_domain.com (invalid because underscores ("\_") are invalid in hostnames. Some DNS records like SRV use them, but not A/AAAA records.)}

\noindent{{\em Test 5:} example-.com (invalid because labels can't start or end with -)}

\noindent{{\em Test 6:} example..com (invalid because of the empty label between dots)}

\noindent{{\em Test 7:} .example.com (invalid because leading dot is not allowed in DNS queries).}


\noindent{{\em Test 8}: example.com.. (invalid because only a single trailing dot is valid to represent the root).}

\noindent{{\em Test 9}: münich.com (invalid because it contains Unicode ü; must be encoded as xn--mnich-kva.com for proper DNS resolution.)}

\noindent{{\em Test 10:} exam ple.com (invalid because spaces are never allowed in DNS names).}

\noindent{{\em Test 11:} example..com (invalid because the root label (empty string) is only valid at the end).
Note: this is the same as test 6 but is generated for a different reason.}


Note how these constraints and tests are more unusual than those a human tester may produce with their (imperfect) recollection of the nuances of the DNS specification.  Note also that extremal testing of inputs finesses the oracle problem~\cite{oracleproblem} of determining what the correct answer should be of the resolution, because the answer for all extremal tests should be to reject because the input invalid.

In a third prompt, we asked GPT-4o for code to run these tests on a DNS implementation.  GPT-4o produced code to test python's `dnspython' library but all tests passed.  
This is the only case where a third prompt was useful; in 
all other cases we had to manually set up test harnesses. 


We did further manual testing using these extremal tests on four widely-used authoritative DNS implementations: BIND, NSD, Knot DNS, and PowerDNS that have been extensively tested ~\cite{kakarla2022scale, kakarla2023oracle} using other methods.
We created a simple zone file for each test case where the invalid domain name is used for an $A$ (IPv4) record.
Each zone file was first submitted to the respective zone file preprocessor and then deployed on actual DNS server instances to observe runtime behavior. For all zone files except Test 3 which contains a domain name with an illegal character (!), all four servers uniformly rejected the zone, either by failing to load the zone or by refusing to answer queries associated with the malformed names.

However, for Test 3, there was a divergence in behavior. Knot DNS rejects the zone at the preprocessing stage, while PowerDNS allows the zone to load but returns \texttt{SERVFAIL} to all queries related to the invalid name or its subdomains. In contrast, both BIND and NSD accept the zone without complaint.
RFC 1035 specifies a formal grammar for domain names which explicitly excludes special characters such as `!', `\_', and spaces from host domain names. However, our experimental results suggest that BIND and NSD are permissive in accepting such names, in apparent violation of the standard. We have reported this issue to the BIND and NSD development teams via their public issue trackers and are currently awaiting their responses.




\section{HTTP Extremal Testing}
\label{sec:http}

URI resolution is a critical feature of HTTP that underlies virtually all web applications. Bugs have been found in popular HTTP implementations that threaten service integrity and confidentiality~\cite{aldoseri2022, hamon2013, reynolds2022}
We leveraged LLMs to generate edge cases based on RFC 3986~\cite{urirfc} and regex expressions. We used the LLM-generated test cases to perform differential testing across five open-source HTTP implementations. With minimal user effort, our methodology generated around 400 subtle test cases across five servers -- httpd, Nginx, Caddy, H2O, and Lighttpd -- and enabled the discovery of two new bugs involving improper handling of null bytes and long URI strings


\begin{lrbox}{\codeg}
\begin{lstlisting}[style=PromptStyle]

model = GPT4(
"""
Consider the general conditions on 
the structure of a valid URI based on 
RFC 3986 and RFC 8820. 

Write code that generates an exhaustive and systematic set of test URIs that violate each constraint. 

Keep the authority and other 
components of the uri separate, like 
{'authority': 'example.com', 'path': '/this/is/valid?q=test#section-1'}.
"""
)

output = model.get_chat_result(
    """
    import string 
    import json

    // Implement the following function generate_uri_test_cases(), which returns a list of malformed URIs. 
    
    // Include helper functions to generate invalid components.
    
    // Start with a valid URI and systematically generate invalid cases for each component.

    def generate_uri_test_cases():
        # Get valid and invalid components 
        
        # Helper to construct bad URIs 
        
        def build_test_case(component_name, bad_value, **kwargs):
    """
)
\end{lstlisting}
\end{lrbox}

\begin{figure}[htbp]
\centering
\begin{tikzpicture}
\node[] (impl) at (0,0) { \usebox\codeg };
\end{tikzpicture};
\vspace{-1em}
\caption {\bf Prompt for function generation.}
\label{fig:prompt.2}
\vspace{-\baselineskip}
\end{figure}

To guide the LLM in function generation, we prompted it (\cref{fig:prompt.2}) to consider each component and the constraints at the component level of a URI independently. The produced code separates the logic for each URI component (scheme, host, port, user info, path, query, fragment) into different modules, allowing users to easily understand which constraints are tested by each string. To systematically generate test cases with a single constraint violation, we begin with a valid URI and iteratively apply each violation in isolation to the relevant segment---modifying only that segment while preserving the integrity of the rest. 

We used OpenAI's Deep Research agent and gave it a general prompt asking for edge cases with justification for each case. 
A sample test case returned by the LLM that would be nearly impossible to generate using model-based testing~\cite{kakarla2023oracle, kakarla2022scale} is \texttt{http://example.com/foo\%00bar}, where there is a null byte present in the path segment. For model-based testing to generate a similar test case, there would need to be an explicit check for a null byte that symbolic execution tool like KLEE can explore in the model. 


All servers handle well-formed URI requests in a manner consistent with the specifications. httpd, Nginx, and Lighttpd agreed on all test cases, including those involving abnormal and irregular URIs---both valid and invalid. 

H2O and Caddy both struggled with URIs containing a null byte (\%00), but their responses differed. While Caddy responds with a 500 code (Internal Server Error), H2O responds with a 301 code (Moved Permanently). The more appropriate response is 400 (Bad Request), which is indeed how httpd, Nginx, and Lighttpd respond. RFC 3986~\cite{urirfc}, Section 7.3, says:
"Note, however, that the "\%00" percent-encoding (NUL) may require special handling and should be rejected if the application is not expecting to receive raw data within a component."

We found Caddy attempts to resolve the path but fails to, resulting in a system-level invalid argument error. However, since the issue stems from the malformed nature of the path itself, a more appropriate response would be to return 400 as a client error to signal that further inspection of the request object is necessary. Further investigation is needed to determine why H2O returns 301 when the null byte is present, but the redirect path is precisely the original request URI and causes a redirect loop. Null byte injection is an active exploitation technique used to fool filters in web infrastructure to gain unauthorized access, making these inconsistencies particularly concerning for security purposes. 

Another case in which Caddy did not agree with the other four servers is handling URIs with extremely long path segments. We used a test case with a path of 1008 characters. The other servers responded with a 403 (Forbidden) or 404 code (Not Found), but Caddy responded with 500.  We investigated further, as there is no formal upper-limit on the length of a URI: thus. this discrepancy could result from different servers defining their own implementation limits. 

To examine our hypothesis, we ran a test with the exact same long string used as the query segment rather than the path. Caddy responded with 404, confirming that the discrepancy does not occur in this scenario, which supports the hypothesis that the error handling is due to the underlying file system limits. Both of the Caddy bugs have been acknowledged by developers and fixes are underway. 

More subtly, RFC 7230~\cite{httprfc}, Section 3.1 says: "A server that receives a request-target longer than any URI it wishes to parse MUST respond with a 414 (URI Too Long) status code".  None of the five servers we tested responded with 414 for the long URI we tested. This may be harmless, but is technically a violation of the RFC. 

{\em Completeness versus Ease of Generation:} For some constraints, e.g., a URI should not contain any of a set $S$ of invalid characters, we were concerned that the LLM may pick one element of $S$ and perhaps another element triggers a bug.  To test this, for HTTP, we took the LLM generated constraints and wrote Python to more completely enumerate invalid input including one test case for every member of a forbidden set, followed by testing on the same 5 HTTP implementations. Note that we wrote the Python script 
{\em only} to test the completeness of the LLM generated tests.


This more exhaustive test regimen did not, however, produce any new bugs and required writing 450 LOC while no code was required for the LLM based process. It may be surprising that the LLM generated the null byte extrema from a field of other possible invalid characters.  We conjecture the LLM may be reasoning directly to find characters that can cause bugs, instead of naively trying all invalid characters as our Python code did.

{\em Limitations:} We did not test advanced features, such as multiplexing, stream prioritization, and connection migration. 
Additionally, the use of a static state URI—the root URI—restricted our testing of relative URIs. 

\section{BGP Extremal Testing}
\label{sec:bgp}

We generate extremal tests for BGP route filtering based on attributes local preference, MED, origin, AS-path, and community. We chose route filtering as a starting point because prior bug reports suggest it is a promising area for uncovering discrepancies. We leave testing other BGP features to future work. Using  Prompt 1 in \cref{fig:initial_prompt}, GPT-4o returned validity constraints for routes and route maps and generated 10 invalid test cases. For example, one test contained an invalid prefix (300.0.0.0/24), and another had an invalid action "drop" in a community-list.

\textbf{Simulating Agentic AI:} A natural next step beyond one-shot test generation is to incorporate LLM-powered agents and 
reflection~\cite{agenticLLMreflection} where a "Test Quality Checker" agent responds to a "Test Generator" agent. For BGP, we simulated this
manually
using Prompt 2, also shown in \cref{fig:initial_prompt}. We leave testing for more rounds to future work using automation via agentic AI.

\textbf{Differential Testing:}
We used differential testing across two widely used implementations and a simulator: FRR, GoBGP, and Batfish~\cite{singha2024messi}. 
The first round test case
revealed two discrepancies.

\textit{Negative Local Preference:} Batfish permitted a route with a negative local preference (`-100'), unlike FRR and GoBGP. While the BGP RFC 4271~\cite{bgprfc} does not explicitly forbid negative values, negative values are not even representable in the IETF YANG model. Further, FRR and GoBGP reject negative values; thus, this is a Batfish bug. We have contacted the Batfish developers and are awaiting their response. 

\textit{Invalid AS-path:} This uncovered a confirmed bug in GoBGP—it accepted an invalid AS-path regex (`[') that should have caused a configuration error. This issue was reported and acknowledged by the GoBGP developers.

In response to the second round Prompt 2 shown in \cref{fig:initial_prompt}, the LLM generated two new test cases using feedback from the first round. One such test case was:

\textit{Conjunctive Match Logic:} This test checks whether implementations enforce conjunctive match logic. A route was sent that matched a prefix but failed  a second AS-path regex match condition. GoBGP permitted it; FRR denied it. Simplifying the regex 
from \verb|^65050| to \verb|65050| 
led both to deny. The discrepancy was due to FRR’s implicit deny and GoBGP’s different regex semantics. While this new test case was
not a bug, it did reveal a subtle behavioral difference between FRR and GoBGP. 


\begin{lrbox}{\codeg}
\begin{lstlisting}[style=PromptStyle]
*************** Prompt 1 ***************
Give me some validity constraints for BGP route and route maps. Then Generate 10 test cases in which either route or rmap or both violates one or more of these validity constraints. Do not include any comments in the json output. A Sample output shown below:
[{"test case": 1,
  "description": ...,
  "route": {"prefix":_, "local-pref":_, "med":_, "as-path":_, "origin":_, "community":_},
  "rmap": {"local-pref":_,"med":_,
  "prefix-list": [{"match":_,"action":_},...], "community-list": [{"match":_,"action":_},...], "as-path-list": [{"match":_,"action":_},...], "rmap-action": "permit"/"deny"}, 
  "expected":"permit"/"deny"}, ...
]
*************** Prompt 2 ***************
This is the differential testing results obtained after running the previous test cases. 
<Test Results for all implementations in a table>. 
Based on these results generate 10 additional test cases to maximize the possibility of finding a bug.
\end{lstlisting}
\end{lrbox}

\begin{figure}[H]
\centering
\begin{tikzpicture}
\node[] (impl) at (0,0) { \usebox\codeg };
\end{tikzpicture};
\vspace{-1em}
\caption{LLM Prompts for Route Filtering}
\label{fig:initial_prompt}
\vspace{-\baselineskip}
\end{figure}

\section{Extremal Testing for Software}
\label{sec:software}

While we have discussed extremal tests for protocol implementations, the methodology applies to regular software.  We tried it on a widely used piece of network software.

\textbf{Shortest path algorithms:} Every OSPF router uses Dijsktra's algorithm to calculate shortest paths. We experimented with extending extremal testing to two popular libraries of graph algorithms: 
networkX~\cite{NetworkX} (python, input as adjacency lists) and Boost~\cite{boost} (C++, input as adjacency matrix). We used the same two-stage query, first asking the LLM to generate logical constraints on the input, and then asking it to generate extremal test cases.

For both libraries, the LLM generated the classic invalid case where the graph contains negative edges. But extremal test generation was far more capable. For Boost, the LLM produced the test case:

\begin{verbatim}
[[0,1e308,INF], [1e308,0,1e308], [INF,1e308,0]]
\end{verbatim}

While each edge has a valid weight within the range of C++ doubles, the length of the shortest path from node 1 to node 3 would overflow a double. Boost reported the incorrect value of $MAX\_DOUBLE$ instead of the correct 2.0e138 or throwing an overflow exception. This should change; by contrast, the well known LEDA package~\cite{leda} provides an option where weights are in a custom integer type that do not overflow~\cite{ledatypes}. LEDA also provides a method to check the results of shortest path algorithms~\cite{ledachecking}.  We have contacted the Boost developers and are awaiting a response.

For NetworkX, the LLM test cases demonstrate subtleties in the library’s design. The \texttt{add\_edge()} function allows the weight parameter of some edges to be undefined (in which case the weight is assumed to be 1).  The LLM produced an example where one edge was undefined and the other had weight 5; this input raised no error on creation, but led to an exception during shortest path execution. The LLM provided another extremal case via a code snippet that generates a complete graph with 100k vertices. This case caused the OS to kill the python interpreter process for memory usage. While these extrema can be confirmed from documentation, they are hard to find manually.


\section{Prompting Strategies}
\label{sec:prompting_startegies}

A natural question is whether the \emph{chain-of-thought} style prompting we used when generating extremal test cases is necessary. To answer this, we explored three strategies:

\noindent{\textbf{Two-Stage Prompt (Chain-of-Thought):}}

\noindent{{\em 1.} Tell me some validity constraints for \{Feature\} of \{Protocol\}.}  

\noindent{{\em 2.} Generate a set of invalid test cases for \{Feature\} that violate one or more of the above constraints and can be used to test real \{Protocol\} implementations.}

\noindent{\textbf{One-Stage Prompt with Validity Constraints:} Generate a set of invalid test cases for \{Feature\} of \{Protocol\} that violate one or more validity constraints and can be used to test real implementations.}

\noindent{\textbf{One-Stage Prompt without Constraints:} Generate a set of invalid test cases for \{Feature\} of \{Protocol\} to test real \{Protocol\} implementations.}

~\cref{tab:prompting_strategies} summarizes the number of test cases generated and the number of \emph{bug-triggering} cases discovered by each strategy across four protocols.

Observe that two-stage prompting consistently yields higher-quality test cases, with a better bug-finding rate per test. One-stage prompting with validity constraints sometimes performs comparably (e.g., DNS: 10 tests, 1 buggy case), but is less consistent. In contrast, when constraints are omitted entirely, the LLM appears to produce overly simplistic inputs that fail to uncover bugs. Overall, explicitly asking the LLM to reason about validity constraints before generating test cases appears more effective, at least with the current state of LLM technology.

\begin{table*}[ht]
\centering
\small
\renewcommand{\arraystretch}{1.2}
\begin{tabular}{|C{2.3cm}|C{1.5cm}|C{1.5cm}|C{2.2cm}|C{2.2cm}|C{2.2cm}|C{2.2cm}|}
\hline
\multirow{2}{*}{\textbf{Implementation}} 
& \multicolumn{2}{c|}{\textbf{Two-Stage Prompting}} 
& \multicolumn{2}{c|}{\textbf{One-Stage w/ Validity Constraints}} 
& \multicolumn{2}{c|}{\textbf{One-Stage w/o Validity Constraints}} \\
\cline{2-7}
& \textbf{Total} & \textbf{Buggy} 
& \textbf{Total} & \textbf{Buggy} 
& \textbf{Total} & \textbf{Buggy} \\
\hline
DNS & 12 & 1 & 10 & 1 & 15 & 0 \\
\hline
BGP & 10 & 2 & 5 & 1 & 5 & 0 \\
\hline
HTTP & 20 & 2 & 14 & 0 & 10 & 0 \\
\hline
Shortest Path & 10 & 1 & 8 & 0 & 6 & 0 \\
\hline
\end{tabular}
\caption{Tests generated and bugs found with different prompting strategies}
\label{tab:prompting_strategies}
\end{table*}

\section{Future Directions}
\label{sec:futurework}

Extremal testing can be extended beyond network software:

{\em 1. Preprocessor to Reject Extrema:} Instead of generating extremal tests, can we ask the LLM to generate preprocessing code to reject invalid input? This is not obvious as the LLM-generated code must check for invalid predicates, not merely generate invalid tests. We did a quick experiment for BGP and for shortest paths; when used as a prefilter for GoBGP, the GoBGP results aligned with other implementations (~\cref{sec:bgp}) and all extremal test cases for shortest paths were weeded out.

{\em 2. Agentic AI for further automation:} We have built a preliminary three agent pipeline using 220 lines of Python code following the ChatDev framework for agentic LLMs~\cite{chatdev}. It comprises LLMs simulating three agents: Researcher (reads RFCs to generate protocol extremal conditions), Senior Researcher (provides feedback on results generated by the researcher) and Intern (writes test cases from the researcher’s list of extremal conditions).  Can an agentic AI system iterate through a list of protocol RFCs with no manual prompting (as we have done so far) and also give feedback~\cite{FacebookLLMTesting} to the Researcher agent to ask it for more extensive extremal tests as we did manually for BGP?

{\em 3. Arbitrary Software:} What (arbitrary) software can extremal testing be applied to? For the LLM to generate extremal constraints, however, the software or problem solved should be well known, as we have seen for shortest paths.

{\em 4. Internal Extremal Testing:} The alert reader will have noticed that our paper focuses on extremal inputs, ignoring extremal values for (internal) state variables within a program. For example, consider a state variable $D = A/(B-C)$ where $A$, $B$, and $C$ are inputs.  The technique in this paper would not detect the divide by zero extrema when $B= C$ if both are valid inputs.  Is there a technique that can combine SAT solvers and LLMs that can generate extremal values for all tests in a program?  SAT solvers used naively will not generate extremal values though white box BVA solves a related 
problem~\cite{zhang2015boundary}.

\section{Related Work}
\label{sec:relatedwork}

\textbf{Boundary value analysis (BVA)} is the closest to extremal 
testing~\cite{bvareview, sholeh2021black, ramachandran2003testing, zhang2015boundary}.  However, BVA does not use an LLM to generate boundary conditions and typically focuses only on  range constraints.  A very recent exception is 
\cite{LLMBVA} which uses an LLM to directly generate boundary value tests without our intermediate step of generating constraints, and also requires the code to be input to the LLM.  Further, the evaluation is only for code of a few hundred lines, much smaller than larger code bases like BGP, DNS or HTTP implementations.

\textbf{LLM driven test generation:} LLM based software testing is a vast area~\cite{LLMTestingSurvey}, a subset of LLM based Software Engineering~\cite{FacebookLLMTesting}.  However, existing work~\cite{LLMTestingSurvey} uses LLMs to improve coverage or to improve mutation/fuzz based testing, not for extremal testing.  Further, LLM testing has been applied to many domains like compilers and mobile apps (see Figure 5 in \cite{LLMTestingSurvey}), but not to network software.


\textbf{Fuzz testing} is widely used for software testing in general~\cite{fuzzer1, fuzzer2, peachfuzzer, bohme2016coverage, 10.1145/2714565} and specifically for BGP and DNS implementations (e.g., \cite{forescoute, frr-fuzzer, cambus, nmap, honggfuzz}). Although fuzzers are effective at finding parser bugs, random inputs have only a small probability of finding extremal bugs.

\textbf{Symbolic execution} uses SMT solvers to generate test cases for many execution paths of a program (e.g., \cite{cadar2008klee,godefroid2005dart}). Protocol implementations contain many thousand lines, making symbolic execution infeasible.

\textbf{Model-based testing} uses an abstract model of a system to generate tests~\cite{bozic2018formal}. It has been used for DNS \cite{kakarla2022scale} and BGP~\cite{singha2024messi} but generates valid, not extremal inputs.  Eywa~\cite{kakarla2023oracle} uses an LLM to generate models, not tests.

\section{Conclusions}

The extremal testing methodology we advocate in this paper uses 
chain-of-thought prompting to first ask an LLM for constraints, and then ask it to generate tests that violate constraints.  The methodology finesses the oracle problem in software testing~\cite{oracleproblem} because it focuses on invalid inputs that must (ideally) be rejected.  Like 
Eywa~\cite{kakarla2023oracle}, LLM hallucinations do not affect correctness and may, in fact,
generate interesting test cases.

While the harvest of bugs we found using this methodology was small, extremal testing quickly found new bugs in mature protocol implementations that have been extensively tested using other methods like symbolic~\cite{cadar2008klee}, fuzz~\cite{frr-fuzzer, peachfuzzer, cambus} and model based~\cite{kakarla2022scale, singha2024messi} methods.  Further, we did not test many HTTP and BGP features, and did not use multiple rounds of LLM test generation except for BGP. 

Finally, it is noteworthy that our prompts can almost be templatized with ``BGP'' replacing ``DNS'' or ``HTTP''.  Perhaps the simplicity and lack of tester expertise may make extremal testing an additional weapon in the arsenal of even the least experienced test engineer.

\section{Acknowledgments}

We thank Todd Millstein for his insightful comments comparing Extremal testing to LLM based test generation for software. The work of Rathin Singha and George Varghese was partly supported by NSF Grant 2402958.

\bibliography{main}
\bibliographystyle{abbrv}
\end{document}